\documentclass[runningheads]{llncs}
\usepackage[T1]{fontenc}
\usepackage{cite}
\usepackage{amsmath,amssymb,amsfonts}
\usepackage{algorithmic}
\usepackage{graphicx}
\usepackage{textcomp}
\usepackage[hyphens]{url}
\usepackage{fancyhdr} 
\usepackage{hyperref}
\usepackage{algorithmic}
\usepackage[linesnumbered,lined,ruled,vlined]{algorithm2e}
\usepackage{bm}
\usepackage{pifont}
\usepackage{soul}
\usepackage{bbding}
\usepackage{wasysym}
\usepackage{pifont}
\usepackage{threeparttable}
\usepackage{multirow}
 
\usepackage{marvosym}
\usepackage[table]{xcolor}
\usepackage{booktabs}
\usepackage{array}

\begin{document}
\title{MOCAP: Wafer-Scale-Chip-Oriented Memory-Orchestrated Chunked Pipelining Framework for Prefill-Only LLM Inference}



\author{Zichuan Wang\inst{1}$^*$ \and
Huizheng Wang\inst{1}$^*$ \and Yuheng Xiao\inst{1} \and Haonan Zuo\inst{1} \and Taiquan Wei\inst{1} \and Jinyi Deng\inst{1} \and Chao Li\inst{2} \and
Yang Hu\inst{1}\textsuperscript{\Letter} \and Shouyi Yin\inst{1,3}} 

%

\institute{School of Integrated Circuits, BNRist, Tsinghua University, Beijing, China, 100084 \and
School of Computer Science and Engineering, Shanghai Jiao Tong University, Shanghai, China, 200240 \and
Shanghai Artificial Intelligence Laboratory, Shanghai, China, 200433 \\ \textsuperscript{\Letter}Corresponding author, hu\_yang@tsinghua.edu.cn \\ $^*$ These authors contributed equally}

\maketitle              

\begin{abstract}
 
Large language models (LLMs) are increasingly used in prefill-only workloads, where end-to-end latency is dominated by the prefill phase. For long-context prefill, communication overhead grows with sequence length and quickly becomes a bottleneck on conventional GPU systems, making wafer-scale chips (WSCs) a promising substrate due to their high communication bandwidth and large aggregate compute and memory capacity. A natural way to accelerate prefill is to partition a long input sequence into multiple chunks and execute them in a finer-grained pipeline across devices. However, directly applying this idea to long-context prefill on WSCs remains challenging. First, causal dependency across chunks causes KV cache to accumulate unevenly across pipeline stages, creating severe memory imbalance and limiting the feasible sequence length. Second, later chunks require more attention computation because each chunk depends on preceding chunks, leading to chunk-level latency imbalance.

To address these challenges, we present MOCAP, a memory-orchestrated chunked pipelining framework for prefill-only LLM inference on WSCs. MOCAP introduces Memory-Balanced KV Reallocation (MBKR) to alleviate memory imbalance by redistributing KV cache across pipeline stages, thereby extending the feasible sequence length. It further incorporates Latency-Balanced Chunk Partitioning (LBCP) to balance chunk execution cost under both attention-cost growth and KV reallocation overhead, improving pipeline efficiency. Experimental results show that, compared with GPipe, MOCAP achieves 76.4\% lower end-to-end latency and 3.24$\times$ higher throughput on average. MOCAP also extends the maximum supported sequence length by up to 1.31$\times$ compared with Terapipe.

\end{abstract}

\section{Introduction} \label{sec:introduction}

\begin{figure}[t]
\includegraphics[width=\textwidth]{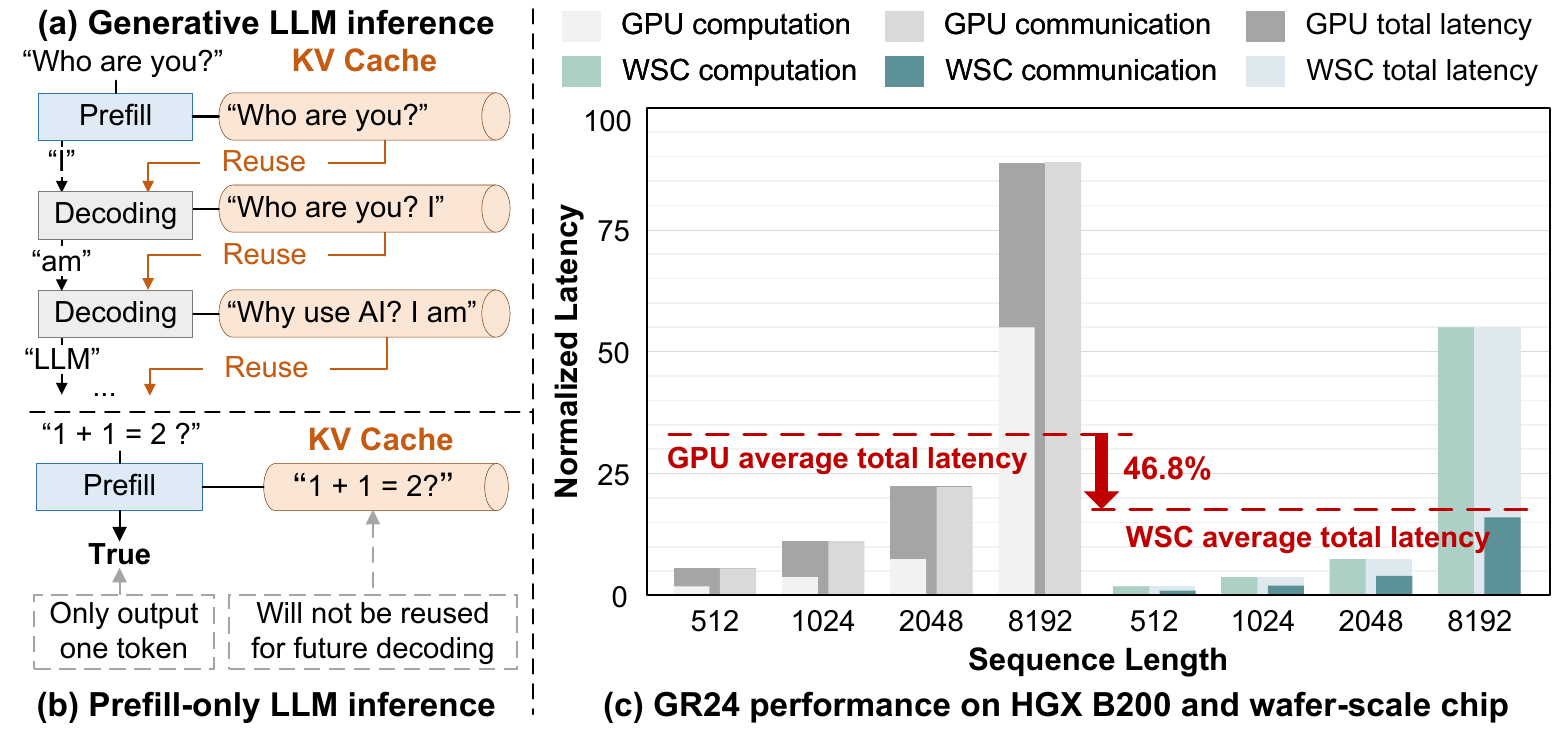}

\caption{(a)(b) LLM inference process. (c) End-to-end latency comparison between HGX B200 GPU systems and wafer-scale chips under equivalent compute and memory capacity on the GR24 \cite{zhai2024actions}.} \label{fig:IN_introduction}

\end{figure}

Large language models (LLMs) have transformed AI across a broad range of applications. Beyond generative appliccations \cite{zhang2025generative}, LLM-style transformer inference is increasingly used in prefill-only scenarios \cite{du2025prefillonly, zhai2024actions}, where the system only needs to process the input context and produce one output token, as shown in Fig.~\ref{fig:IN_introduction} (b). In these workloads, the end-to-end latency is largely determined by the prefill phase, making prefill optimization essential. 

However, since the communication volume in prefill grows proportionally with sequence length, communication becomes the dominant bottleneck on widely deployed GPU systems, as illustrated in Fig.~\ref{fig:IN_introduction} (c). Recently, chip integration has advanced toward wafer-scale chips (WSCs) \cite{hu2024wafer, lie2022cerebras, talpes2022dojo, shih2025sow, pal2021designing, liu2026ouroboros}, which provide much higher compute density, memory capacity, and communication bandwidth. This makes WSCs a promising substrate for communication-intensive prefill. As shown in Fig.~\ref{fig:IN_introduction} (c), the communication advantage of WSCs reduces end-to-end latency by 46.8\% compared with GPU systems. However, existing studies \cite{he2025waferllm, xu2025wsc} mainly focus on general inference, and do not specifically optimize prefill-only execution.

Terapipe \cite{li2021terapipe} proposes a token-wise pipeline for the forward pass of LLM training, which exploits the causal dependency along the token dimension to enable token-wise pipelined execution across devices. Inspired by this idea, prefill can similarly partition the sequence into multiple chunks and execute them as a fine-grained chunked pipeline to reduce pipeline bubbles and improve utilization. 

However, directly extending such a pipeline to prefill remains challenging. First, causal dependency across chunks causes the KV cache to accumulate unevenly across pipeline stages, and the resulting memory pressure quickly becomes a bottleneck, limiting the feasible sequence length. Second, the execution cost of different chunks is inherently imbalanced: later chunks incur higher attention latency due to their longer prefixes, so uniform chunk partitioning creates pipeline bubbles and degrades pipeline efficiency. Therefore, making the chunked prefill pipeline effective requires jointly addressing both KV cache accumulation and chunk-level latency imbalance.

To this end, we present MOCAP, a memory-orchestrated chunked pipelining framework for efficient prefill-only LLM inference on wafer-scale chips. First, MOCAP introduces Memory-Balanced KV Reallocation, which redistributes KV cache across pipeline stages to mitigate memory imbalance and relieve KV cache buildup at earlier stages, thereby enabling longer sequence execution under limited memory. Second, MOCAP incorporates Latency-Balanced Chunk Partitioning, which adaptively partitions the input sequence into non-uniform chunks to balance KV cache reallocation overhead and reduce pipeline bubbles caused by the increasing attention latency of later chunks. Together, MOCAP transforms the chunked pipeline into a scalable and efficient execution paradigm for prefill-only LLM inference on WSCs.

Our contributions are summarized as follows:

(1) We propose MOCAP, a memory-orchestrated chunked pipelining framework for prefill-only LLM inference on wafer-scale chips. \textbf{To the best of our knowledge, this is the first work to systematically optimize prefill-only LLM inference on WSCs}.

(2) We develop Memory-Balanced KV Reallocation, which mitigates stage-wise KV cache accumulation by redistributing KV cache across pipeline stages, improving memory utilization and enabling longer sequence execution.

(3) We design Latency-Balanced Chunk Partitioning, which balances KV cache reallocation overhead and compute imbalance through non-uniform partitioning.

(4) On average, MOCAP reduces end-to-end latency by 76.4\% and improves throughput by 3.24$\times$ over GPipe. It also extends the maximum supported sequence length by up to 1.31$\times$ compared with Terapipe.

\begin{figure}[t]
\includegraphics[width=\textwidth]{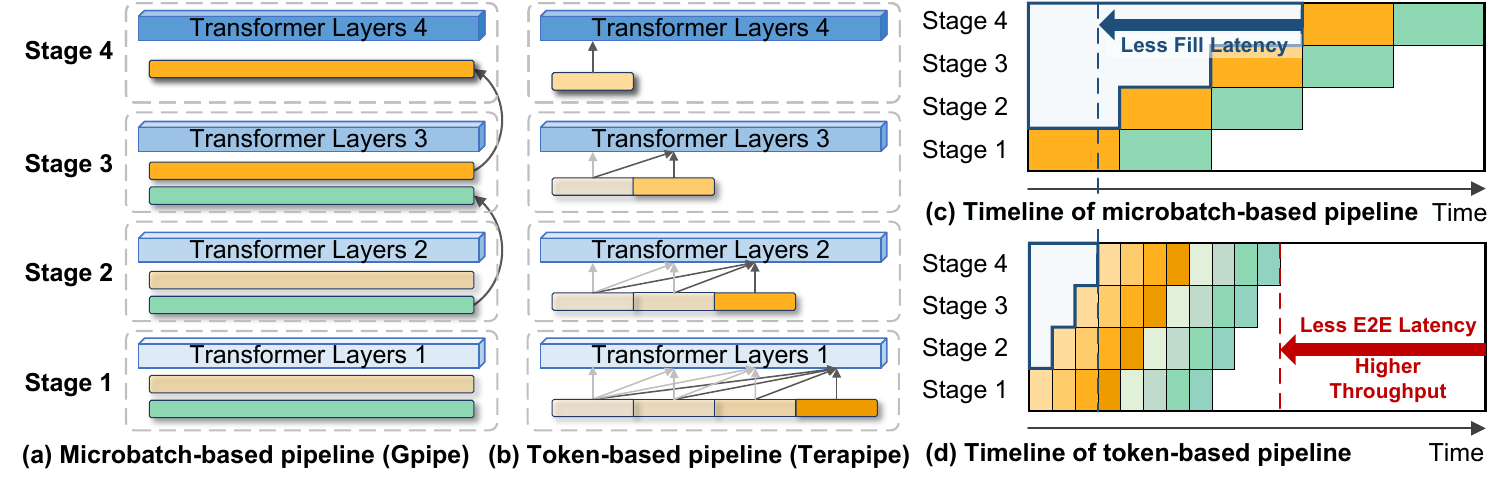}

\caption{Overview of microbatch-based pipeline and token-based pipeline.} \label{fig:BW_PP}

\end{figure}

\section{Background}\label{sec:background}

\subsection{Prefill-Only LLM Inference}\label{subsec:llm inference system}

Prefill-only LLM inference \cite{du2025prefillonly, zhai2024actions} represents an emerging class of workloads where the model processes the entire input sequence to produce only one token, without long autoregressive decoding. Unlike generative inference, which repeatedly reuses the KV cache during decoding, as shown in Fig.~\ref{fig:IN_introduction} (a), prefill-only inference does not require the generated KV cache to be reused for future long decoding. Because the output is produced immediately after prefilling, the overall service latency is largely determined by the prefill phase.

This execution pattern leads to several distinctive system characteristics. First, although the KV cache is generated during prefilling, it does not need to be persistently stored for future decoding reuse, fundamentally changing the role of KV-cache management. Second, practical prefill-only workloads often involve long input contexts of up to $10^5$ \cite{zhai2024actions}, so the system bottleneck shifts from iterative decoding efficiency to efficient processing of long sequences.

\subsection{Chunked Pipeline}\label{subsec:chunked pipeline}

Terapipe \cite{li2021terapipe} introduces chunked pipeline to alleviate the coarse granularity of microbatch-based pipeline parallelism. Unlike GPipe \cite{huang2019gpipe}, which pipelines microbatches across stages as shown in Fig.~\ref{fig:BW_PP}(a), Terapipe pipelines token chunks across stages (Fig.~\ref{fig:BW_PP}(b)), thereby shortening pipeline fill time and improving throughput, as illustrated in Fig.~\ref{fig:BW_PP}(c)(d). Inspired by this idea, prefill can likewise partition the input sequence into chunks and execute them in a chunked pipeline.

Chunked pipeline induces a cross-chunk dependency through causal attention. As shown in Fig.~\ref{fig:BW_PP}(b), each chunk must attend to the key-value tensors generated by all preceding chunks. This dependency has two implications: 1) KV caches from earlier chunks must remain live for subsequent computation, causing progressive accumulation over the pipeline lifetime. 2) Later chunks face longer prefixes and therefore higher attention latency, which can manifest as pipeline bubbles.

\begin{figure}[t]
\includegraphics[width=\textwidth]{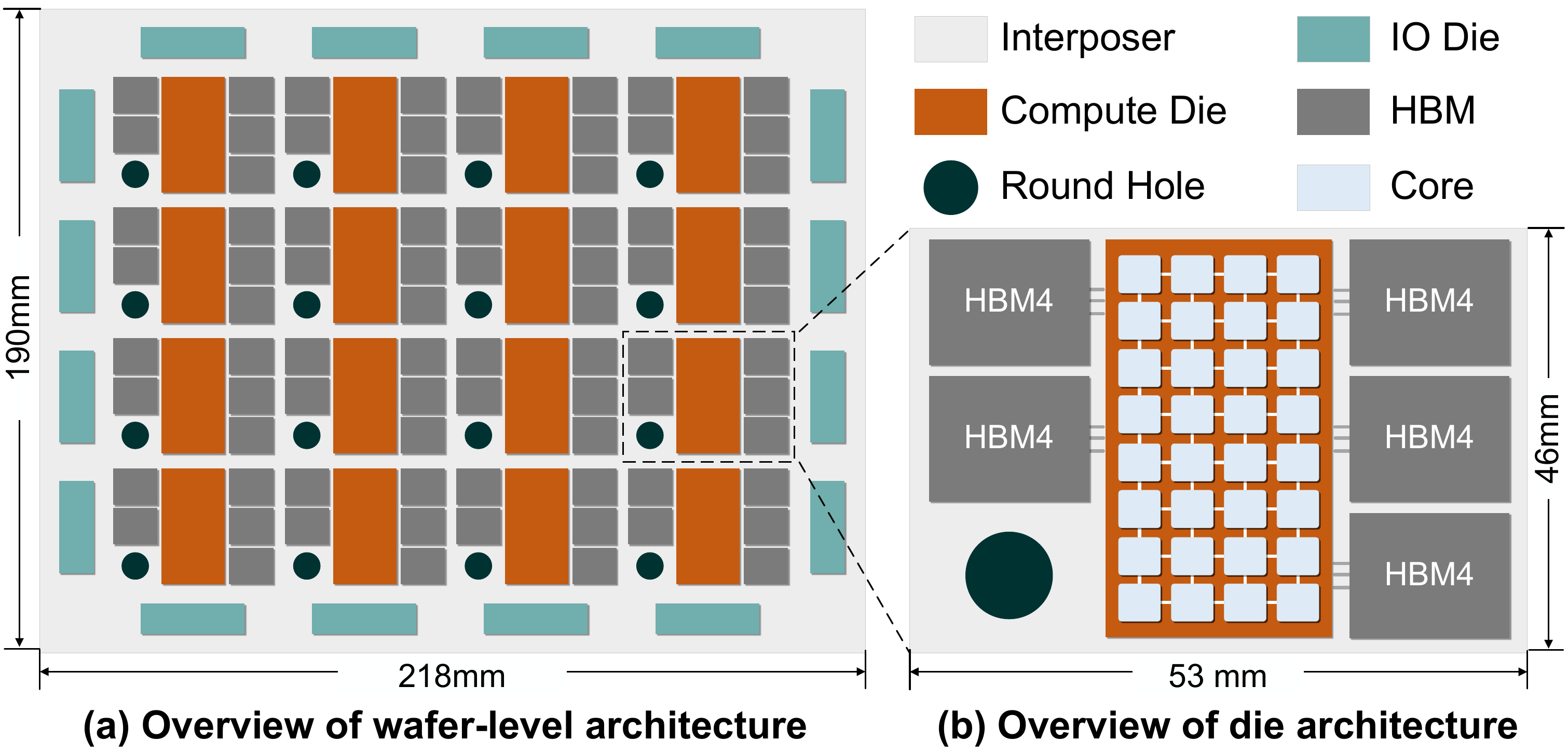}

\caption{Overview of wafer-scale chip architecture.} \label{fig:Background}

\end{figure}

\subsection{Wafer-Scale Chips}\label{subsec:wafer-scale chip}

Wafer-scale chips (WSCs) \cite{talpes2022dojo, lie2022cerebras, shih2025sow, hu2024wafer} extend chip integration beyond the conventional reticle limit through advanced packaging technologies. As illustrated in Fig.~\ref{fig:Background} (a), WSCs consist of an interposer and multiple known-good dies (KGDs). The interposer provides dense on-package interconnects, in which die-to-die spacing is minimized to satisfy signal integrity constraints and maximize communication bandwidth. To preserve ultra-short inter-die links while maintaining scalable routing complexity, wafer-scale chips commonly adopt a nearest-neighbor 2D mesh topology, where each die physically connects to adjacent neighbors. Fig.~\ref{fig:Background} (b) further shows a representative die architecture, where each compute die contains an array of compute cores equipped with distributed SRAM for local buffering, while HBM stacks are placed around the compute die to provide large-capacity off-die storage and high memory access bandwidth.

A key architectural distinction between wafer-scale chips and conventional GPU systems lies in the communication fabric. \textbf{1)} Wafer-scale chips provide up to 10 TB/s die-to-die bandwidth, which is sufficient to largely mask KV-cache transfer overhead compared with the 900 GB/s NVLink \cite{nvidia_nvl72} bandwidth in widely deployed GPU systems. \textbf{2)} The same communication interface can also be used by edge IO dies, extending the high-bandwidth domain beyond a local package region. \textbf{3)} D2D bandwidth becomes comparable to HBM bandwidth, making cross-die data movement a practical option.

\section{Motivation}\label{sec:motivation}

\begin{figure}[t]
\includegraphics[width=\textwidth]{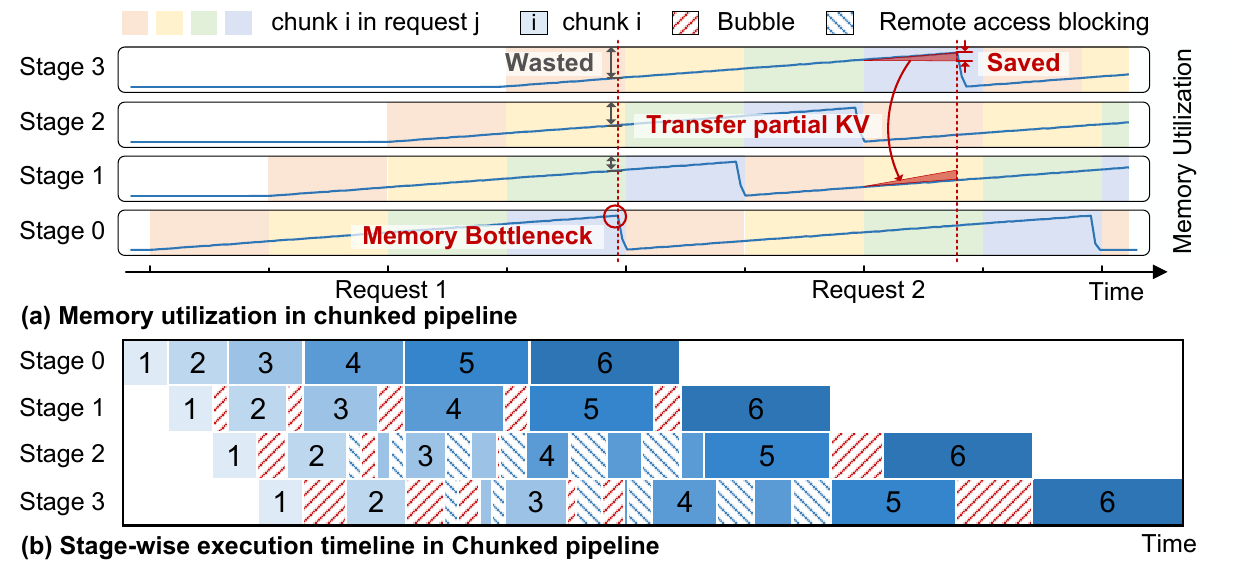}

\caption{Stage-wise memory and execution latency imbalance introduced by chunked pipeline.} \label{fig:MO_all}

\end{figure}

\subsection{Memory Imbalance Limits Feasible Sequence Length}\label{subsec:storage_imbalance}

\textbf{Challenge: As the sequence length grows, stage-wise KV-cache accumulation becomes increasingly imbalanced, eventually limiting the feasible sequence length.} As illustrated in Sec.~\ref{subsec:chunked pipeline}, earlier stages start accumulating KV cache earlier and keep it for longer, while later stages experience a much shorter accumulation window. This asymmetric KV-cache lifetime causes the peak memory occupancy to concentrate on the earliest stages. As illustrated in Fig.~\ref{fig:MO_all} (a), the system is effectively bounded by the earliest-stage memory bottleneck: once the earliest stage reaches its memory limit, the execution can no longer scale to longer sequences, even though substantial memory capacity on later stages remains unused. In other words, the feasible sequence length is constrained by the peak per-stage occupancy rather than by the aggregate memory capacity of the system, leaving a large fraction of memory stranded on non-bottleneck stages.

\textbf{Insight: Such skewed memory utilization creates an opportunity to extend sequence headroom through stage-aware KV-cache reallocation.} Although KV-cache accumulation is highly skewed, the memory pressure is not simultaneously saturated across all stages. In particular, when the earliest stage approaches its memory limit, later stages may still have substantial available space. As highlighted by the \textit{wasted} and \textit{saved} regions in Fig.~\ref{fig:MO_all} (a), transferring part of the accumulated KV cache away from the bottleneck stage can reduce the peak memory occupancy and convert otherwise wasted memory on other stages into effective sequence headroom. Motivated by this opportunity, we propose Memory-Balanced KV Reallocation (MBKR) to mitigate memory imbalance and extend the feasible sequence length without requiring additional hardware memory.

\subsection{Latency Imbalance Limits Compute Utilization}\label{subsec:execution_imbalance}

\textbf{Challenge: Chunk latency becomes increasingly imbalanced as execution progresses, which limits the compute utilization of the chunked pipeline.} The latency of each chunk consists of two major components: the attention latency and the GEMM latency, where the latter includes the QKV projection, the output projection, and the FFN computation. For the $i$-th chunk with length $c_i$, these two components can be approximated as

\begin{equation}
T_{i,\text{attn}} \propto c_i (p_i + c_i), p_i = \sum_{j=1}^{i-1} c_j \qquad
T_{i,\text{gemm}} \propto c_i,
\label{eq:chunk_breakdown}
\end{equation}

where $p_i$ denotes the prefix length preceding the $i$-th chunk. This decomposition reveals the source of imbalance: as execution proceeds, the prefix length grows monotonically, causing later chunks to incur increasingly higher attention latency, while the GEMM latency only scales with the chunk size itself. As shown in Fig.~\ref{fig:MO_all} (b), uniformly partitioned chunks therefore exhibit increasingly imbalanced execution time, which manifests as visible pipeline bubbles. Moreover, once KV cache is reallocated across stages, some stages must fetch KV data from remote stages during execution. These remote fetches do not merely add communication overhead; the accessed stage must spend time serving the request, which can delay its own local computation and thereby extend the critical path. As shown by the blue shaded blocks in Fig.~\ref{fig:MO_all}(b), this additional blocking further aggravates chunk-level latency imbalance.

\textbf{Insight: This inefficiency can be mitigated by reshaping chunk boundaries to compensate for both prefix-induced attention growth and reallocation-induced blocking.} Since later chunks are inherently more expensive, assigning them the same size as earlier chunks is suboptimal. Instead, later chunks can be made smaller while earlier chunks take relatively larger portions of the sequence, so that the reduction in chunk size offsets the growth of attention latency as well as the additional delay caused by remote KV accesses. As suggested by the reduced bubbles and blocking intervals in Fig.~\ref{fig:MO_all}(b), such latency-aware partitioning can better balance chunk execution time, shorten the critical path, and improve effective overlap across stages. To address this issue, we propose Latency-Balanced Chunk Partitioning (LBCP) to mitigate latency imbalance and improve compute utilization in chunked pipeline execution.

\section{MOCAP Framework}\label{sec:framework}

\subsection{Memory-Balanced KV Reallocation}\label{sec:tech 2}

\begin{figure}[t]
\includegraphics[width=\textwidth]{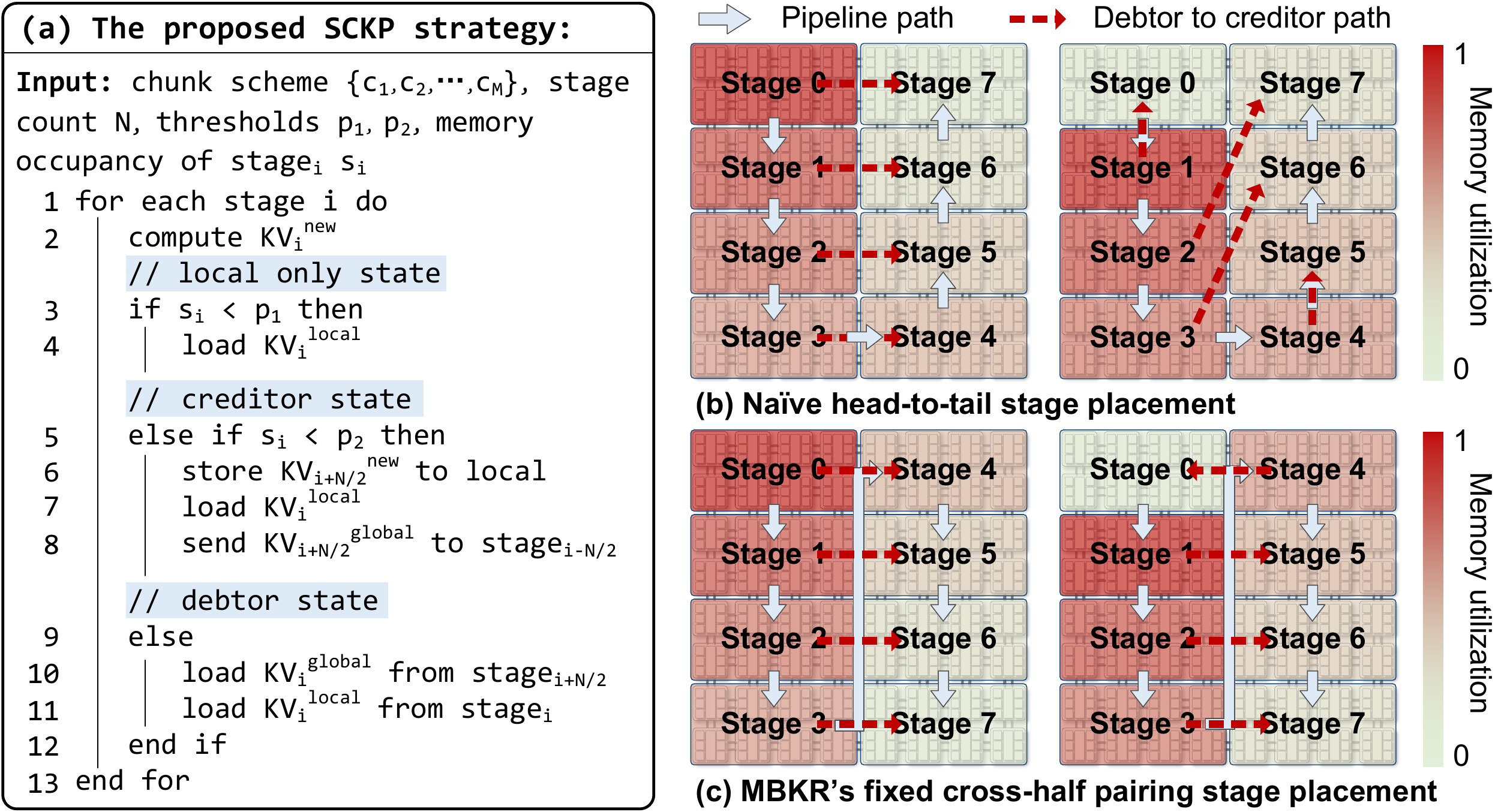}

\caption{Overview of memory-balanced KV reallocation.} \label{fig:FW_MBKR}

\end{figure}

As discussed in Sec.~\ref{subsec:storage_imbalance}, the chunked pipeline creates a predictable stage-wise KV-cache imbalance. Exploiting this opportunity faces two challenges: (1) identifying a stage-pairing strategy that maximizes usable memory capacity, and (2) minimizing the communication cost introduced by KV-cache transfers. Therefore, an effective reallocation scheme must jointly optimize memory utilization and transfer overhead.

To this end, we propose \textit{Memory-Balanced KV Reallocation} (MBKR). MBKR first identifies a stage placement strategy that maximizes memory utilization, and then applies a runtime reallocation policy. To describe this interaction, we refer to stages under excessive memory pressure as \textit{debtor stages}, since they need to spill part of their KV cache, and stages with sufficient residual capacity as \textit{creditor stages}, since they host spilled KV data and serve remote KV requests.

For stage placement, prior work on LLM training \cite{wang2026watos, kim2023bpipe} adopts a head-to-tail pairing strategy, which pairs the stage with the highest memory pressure with the stage that has the lowest memory pressure, as shown in Fig.~\ref{fig:FW_MBKR} (b, left). Such a placement is effective when the memory-pressure pattern remains stable. However, in chunked pipeline execution, the imbalance pattern changes during request transitions, causing the pairing to become mismatched. In particular, when the hotspot shifts to Stage 3, its original paired stage, Stage 4, differs from it by only one chunk in memory occupancy, which means this head-to-tail strategy provides at most half a chunk of additional usable space. Fig.~\ref{fig:FW_MBKR}(b, right) illustrates a natural remedy to dynamically rematch debtor and creditor stages according to instantaneous memory usage, so that the most heavily loaded stage is always paired with the least loaded one. However, such dynamic rematching would require previously spilled KV cache to be migrated again whenever the pairing changes, introducing substantial extra communication and severe link contention.

To avoid this overhead, MBKR adopts a fixed cross-half pairing, i.e., Stage $i$ is paired with Stage $i+
\frac{N}{2}$, and allows each paired stage to dynamically assume debtor or creditor roles according to runtime memory pressure. As shown in Fig.~\ref{fig:FW_MBKR} (c, left), this pairing is effective in the steady phase because the two paired stages consistently differ by $\frac{N}{2}$ chunks in memory pressure, thereby preserving the maximum spill capacity. More importantly, the same relation remains valid in the switching phase. As shown in Fig.~\ref{fig:FW_MBKR} (c, right), although the memory hotspot shifts across stages during request transitions, each stage is still paired with another stage whose pressure differs by $\frac{N}{2}$ chunks, so the fixed pairing continues to provide sufficient spill space without online rematching. In addition, MBKR places stage $i$ adjacent to stage $i+\frac{N}{2}$, so that KV-cache transfers between paired stages traverse the shortest path. This minimizes communication overhead and reduces link contention caused by spill and remote-access traffic. In this way, MBKR simultaneously maximizes usable spill space and minimizes transfer cost under a fixed pairing strategy.

Fig.~\ref{fig:FW_MBKR} (a) further illustrates the runtime policy of MBKR in an SPMD form. At the beginning of the chunked pipeline, every stage computes the newly generated KV cache (lines 1-2). If $s_i < p_1$, the stage remains local-only and loads all required KV cache locally (lines 3-4). If $p_1 \le s_i < p_2$, the stage acts as a creditor: it stores newly spilled KV cache from its paired stage while still serving its own local accesses, and sends the globally visible remote portion to the corresponding debtor stage when needed (lines 5-8). Otherwise, when $s_i \ge p_2$, the stage acts as a debtor and splits its KV accesses into a remote portion loaded from stage $i+\frac{N}{2}$ and a local portion loaded from itself (lines 9-11). In this way, MBKR turns a fixed cross-half pairing into a threshold-driven runtime KV-reallocation policy, without incurring the migration cost of online rematching.

\begin{algorithm}[t]
\begingroup
\linespread{1.0}\selectfont
\small
\setlength{\baselineskip}{0.96\baselineskip}
\caption{\small{Latency-balanced Chunk Partitioning Scheme}}
\label{alg:LCPS}
\small
\KwIn{
    \#Stage $N$, \#Chunk $M$, sequence length $S$
}
\KwOut{Optimal batch size $B^*$, chunk slice scheme $ss^*$, E2E latency $T^*$}

\For{$m \gets M$ \KwTo $1$}{
    \For{$s \gets S-1$ \KwTo $0$}{
        \For{$k \gets 1$ \KwTo $S-s-(M-m)$}{
            $t \gets \textsc{EvaluateChunk}(k,s)$\;
            $t_{max}' \gets \max(t_{\max}[m+1][s+k], t)$,
            $t_{\sum}' \gets t_{\sum}[m+1][s+k] + t$\;
            \If{$t_{\sum}' + (N-1)t_{max}' < t_{\sum}[m][s] + (N-1)t_{\max}[m][s]$}{
                $t_{\max}[m][s] \gets t_{max}'$, $t_{\sum}[m][s] \gets t_{\sum}'$,
                $ss[m][s] \gets k$\;
            }
        }
    }
}
\While{$Temp > Temp_{\min}$}{
    \For{$r \gets 1$ \KwTo $R$}{
        $ss' \gets \textsc{Perturb}(ss)$\;
        $(B', T'_{\mathrm{pre}}) \gets \textsc{EvaluatePrefill}(ss'),~T' \gets \textsc{EvaluateE2E}(B', T'_{\mathrm{pre}})$\;
        \If{$T' < T$ \textbf{or} $\textsc{Rand}(0,1) < \exp(-(T'-T)/Temp)$}{
            $ss \gets ss'$, $B \gets B'$, $T \gets T'$\;
        }
        \If{$T < T^*$}{
            $ss^* \gets ss$, $B^* \gets B$, $T^* \gets T$\;
        }
    }
    $Temp \gets \alpha \cdot Temp$\;
}
\endgroup
\end{algorithm}

\subsection{Latency-Balanced Chunk Partitioning} \label{subsec:FR_LBCP}

As illustrated in Sec.~\ref{subsec:chunked pipeline}, chunked pipeline suffers from two sources of chunk-level latency imbalance: the monotonic growth of attention latency with prefix length, and the additional KV reallocation overhead introduced by remote KV accesses. Together, these factors make chunk latency depend not only on the chunk size, but also on the chunk position and cross-stage KV reallocation. Under this coupled effect, end-to-end latency depends on the full chunk slice scheme and therefore no longer admits the local optimal substructure required by exact dynamic programming.

To address this challenge, we propose \textit{Latency-Balanced Chunk Partitioning} (LBCP). LBCP first derives a compute-balanced chunk slicing scheme as an initialization, and then refines it under the full MBKR-enabled execution model, allowing it to efficiently optimize chunk boundaries for end-to-end latency.

In the first stage, LBCP uses dynamic programming to derive a compute-balanced chunk slice scheme, which serves as the initialization for the subsequent search for the partition that minimizes end-to-end latency (lines 1-7 in Alg.~\ref{alg:LCPS}). The loops enumerate the chunk index $m$, the current position $s$, and the candidate chunk size $k$ (lines 1-3). \textsc{EvaluateChunk} returns the compute latency $t$ (line 4), which is combined with the suffix state at $(m+1, s+k)$ to form $t'_{\max}$ and $t'_{\sum}$ (line 5). The algorithm selects the candidate that minimizes $t'_{\sum} + (N-1)t'_{\max}$ and records the corresponding $k$ in the chunk slice scheme $ss$ (lines 6-7). This proxy yields a compute-balanced partition that provides a robust initialization for the second-stage search. Since it captures only deterministic compute cost, varying the batch size does not affect the relative ranking of candidate partitions.

In the second stage, LBCP refines the partition via simulated annealing under the full MBKR-enabled execution model (lines 8-16 in Alg.~\ref{alg:LCPS}). Each iteration perturbs one chunk boundary while preserving $S$ and $M$ (lines 8--10). \textsc{EvaluatePrefill} returns the maximum feasible batch size $B'$ and prefill latency $T'_{\mathrm{pre}}$. \textsc{EvaluateE2E} then computes the corresponding end-to-end latency $T'$ based on $B'$ and $T'_{\mathrm{pre}}$ (line 11). Better candidates are accepted directly, while worse ones are accepted with a temperature-controlled probability (lines 12-13). The algorithm updates the best solution and gradually cools the temperature (lines 14-16). This refinement captures the coupled effect of compute growth and KV reallocation overhead ignored by the first-stage proxy, and further improves the partition for end-to-end latency. In this way, LBCP efficiently optimizes chunk boundaries for end-to-end latency under both compute growth and KV reallocation overhead.

\begin{figure}[t]
\includegraphics[width=\textwidth]{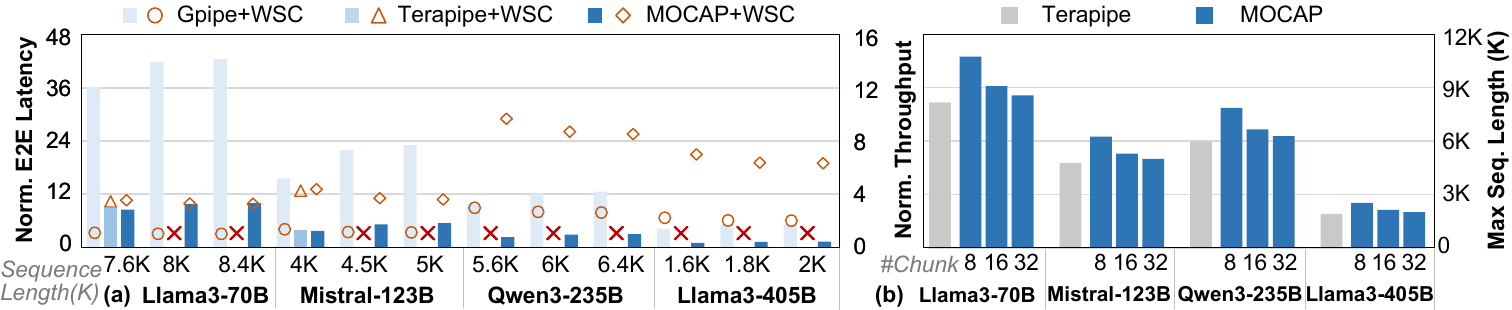}

\caption{(a) Performance of GPipe, Terapipe and MOCAP on WSCs. Red crossed marks indicate that the corresponding prefill jobs fail to run due to out-of-memory errors. (b) Maximum sequence length of Terapipe and MOCAP across different models and number of chunks. } 

\label{fig:EV_WSC}
\end{figure}

\section{Evaluation}\label{sec:evaluation}

\subsection{Experimental Setup}

\textbf{Hardware Configuration}: We configure each die in the WSC to be equivalent to an NVIDIA Blackwell GPU~\cite{nvidia_nvl72} capable of 4.5 PFLOPS, equipped with 180GB HBM featuring 7.7~TB/s access bandwidth. According to SoW-X~\cite{shih2025sow}, D2D bandwidth and cross-wafer bandwidth are both set at 5~TB/s. A 4×4 WSC configuration delivers 72 PFLOPS and 2.8TB memory capacity.

\textbf{Workload}: Given that model size is a key factor influencing performance, we select a series of models from 70B to 405B for our experiments, including Llama3-70B~\cite{Llama3_70B}, Mistral-123B~\cite{Mistral_123B}, Qwen3-235B~\cite{Qwen3_235B}, and Llama3-405B~\cite{llama3_405b}.

\textbf{Methodology}:
To evaluate our proposed strategies, we develop a custom event-driven simulator built upon ASTRA-sim 2.0 \cite{won2023astra}. We adopt end-to-end (E2E) latency and throughput as our primary evaluation metrics. E2E latency denotes the average time elapsed per request from arrival to completion, encompassing queuing delays and prefill execution latency. Throughput is defined as the total number of requests processed per second (req/s).

\subsection{Performance on Wafer-scale Chips} \label{subsec:EV_WSC}

\textbf{Performance compared with GPipe and Terapipe.} We first evaluate MOCAP against two baselines, GPipe \cite{huang2019gpipe} and Terapipe \cite{li2021terapipe}. GPipe serves as a conventional microbatch-based pipeline, while Terapipe represents a chunked pipeline. We perform this comparison on a $4\times4$ WSC across four representative LLMs and vary the sequence length. This setting allows us to jointly evaluate MOCAP under diverse model scales and attention characteristics.

Fig.~\ref{fig:EV_WSC}(a) presents the performance comparison among MOCAP, GPipe, and Terapipe on the WSC across multiple LLM models and sequence lengths. MOCAP consistently achieves the lowest normalized end-to-end latency and the highest normalized throughput across all evaluated settings. On average, compared with GPipe, MOCAP reduces end-to-end latency by 76.4\% and improves throughput by 3.24$\times$. These gains mainly arise from two factors: 1) KV reallocation recovers memory headroom under severe KV accumulation, allowing chunked pipelining to remain effective in the long-context regime; and 2) imbalance-aware partitioning reduces pipeline bubbles and mitigates the performance impact of the additional reallocation overhead. Furthermore, the results reveal two additional insights: 1) MOCAP’s gain is shaped not only by nominal model size, but also by model architecture. Qwen3-235B is an MoE model, whose sparse activation lowers effective per-token compute, while Llama3-405B adopts a larger GQA ratio, which reduces KV-cache footprint and weakens the memory bottleneck from KV accumulation. 2) MOCAP’s relative gain decreases with sequence length, indicating that the increasing KV reallocation overhead gradually offsets a larger fraction of the benefit from chunked pipeline.

\textbf{Maximum sequence length compared with Terapipe.} We further evaluate the maximum feasible sequence length of MOCAP relative to Terapipe, and observe an average improvement of up to 1.31$\times$, as shown in Fig.~\ref{fig:EV_WSC} (b). We also find that using fewer chunks generally allows longer feasible sequences. Under a fixed sequence length, fewer chunks imply a larger chunk size, which enlarges the memory imbalance across stages: earlier stages accumulate KV cache more aggressively, while later stages leave more memory space unused. This stronger stage-wise imbalance creates more reallocation headroom for MBKR and thus supports a larger amount of spilled KV cache. However, larger chunks increase pipeline fill time, reducing pipeline efficiency. This reveals a fundamental tradeoff in chunked pipeline: fewer chunks favor longer feasible sequences, while more chunks favor higher pipeline efficiency.

\section{Related Work}\label{sec:related work}



Prior studies \cite{xu2025wsc,yang2025pd,wang2026watos,tang2026moentwine,he2025waferllm,rashidi2024fred,hu2024wafer,wang2026temp,wang2024tmac,xu2026face,li2026rethermal,wei2025spatial,wang2025designing,liu2026ouroboros,fang2024palm} have explored architecture--system co-design for wafer-scale AI systems across LLM training and inference. Representative efforts such as WSC-LLM \cite{xu2025wsc} jointly optimize LLM inference serving and wafer-scale architectures, whereas WATOS \cite{wang2026watos} co-explores LLM recomputation strategies and wafer-scale architectures for training. Nevertheless, these studies focus on general serving or training scenarios, rather than the unique memory-access and scheduling challenges of prefill-only inference.

Beyond WSCs, prior LLM accelerators \cite{wang2024sofa,wang2025mcbp,wang2026pade,lu2021sanger,wang2026lapa,wang2026bitstopper,yue2024exploiting,wang2025beta} improve Transformer inference at the operator or standalone-accelerator scale through attention sparsity \cite{wang2024sofa,wang2025beta}, bit-level compute--memory co-optimization \cite{wang2025mcbp,wang2026pade}, and data compression \cite{moon2025t,lee2025clat}. Nevertheless, they do not address orchestration and scheduling for multi-die parallel execution. In contrast, MOCAP targets prefill-only LLM inference on WSCs with a memory-orchestrated chunked pipelining framework.


\section{Conclusion}\label{sec:conclusion}

This paper presents MOCAP, a memory-orchestrated chunked pipelining framework for prefill-only LLM inference on wafer-scale chips. By combining MBKR and LBCP, MOCAP jointly mitigates memory imbalance and compute imbalance in chunked pipeline execution, thereby extending the feasible sequence length and improving compute utilization. Compared with GPipe, MOCAP achieves 76.4\% lower end-to-end latency and 3.24$\times$ higher throughput on average. Compared with Terapipe, MOCAP extends the maximum supported sequence length by up to 1.31$\times$.

\section*{Acknowledgments}
This work was supported in part by the National Science and Technology Major Project under Grant 2022ZD0115200; by NSFC under Grant 92464302, Grant 62125403, Grant U24B20164, Grant U24A20234 and Grant 62502255; in part by the Beijing S\&T Project Z251100008425010; Shanghai Municipal Science and Technology Major project; the Natural Science Foundation of Jiangsu Province Basic Research Program under Grant BK20243042; the BNRist; and the Beijing Advanced Innovation Center for Integrated Circuits.

\section*{Declaration on the Use of Artificial Intelligence}
Artificial intelligence tools, if used, were used only for language organization and figure/table polishing, and did not participate in the generation of the core ideas or technical content.

\bibliographystyle{splncs04}
\bibliography{main}

\end{document}